\documentclass[aps,prl,twocolumn,notitlepage,showpacs]{revtex4-1}

\usepackage{amsmath}
\usepackage{amssymb}
\usepackage{graphicx}
\usepackage{epstopdf}
\usepackage[colorlinks=true]{hyperref}

\begin{document}
\title{Magnetic Domain Walls as Hosts of Spin Superfluids and Generators of Skyrmions}

\author{Se Kwon Kim}
\author{Yaroslav Tserkovnyak}
\affiliation{Department of Physics and Astronomy, University of California, Los Angeles, California 90095, USA}

\date{\today}

\begin{abstract}
A domain wall in a magnet with easy-axis anisotropy is shown to harbor spin superfluid associated with its spontaneous breaking of the U(1) spin-rotational symmetry. The spin superfluid is shown to have several topological properties, which are absent in conventional superfluids. First, the associated phase slips create and destroy skyrmions to obey the conservation of the total skyrmion charge, which allows us to use a domain wall as a generator and detector of skyrmions. Secondly, the domain wall engenders the emergent magnetic flux for magnons along its length, which are proportional to the spin supercurrent flowing through it, and thereby provides a way to manipulate magnons. Thirdly, the spin supercurrent can be driven by the magnon current traveling across it owing to the spin transfer between the domain wall and magnons, leading to the magnonic manipulation of the spin superfluid. The theory for superfluid spin transport within the domain wall is confirmed by numerical simulations.
\end{abstract}


\maketitle

\emph{Introduction.}|Under normal conditions, particles in a fluid move against the friction force caused by, e.g., scattering with the vessel. In some extreme circumstances, certain fluids become superfluids that support particle flow with no resistance, which is exemplified by liquid $^4$He at temperatures below 2K~\cite{KapitzaNature1938, *AllenNature1938}. Conventional superfluidity is characterized by the spontaneously broken U(1) symmetry associated with the phase of the macroscopic quantum wave function. A conserved quantity corresponding to the U(1) symmetry is the particle number, or equivalently mass, and it is this mass supercurrent that is carried by the gradient of the phase.

Analogously, materials with the U(1) spin-rotational symmetry can support superfluid spin transport if the symmetry is spontaneously broken by the ground states~\cite{SoninAP2010, *KonigPRL2001, *ChenPRB2014-2, *ChenPRB2014-3, *ChenPRB2014, TakeiPRL2014, *TakeiPRL2015}. Easy-plane magnets thus can realize spin superfluid by choosing an arbitrary direction in the easy plane and thereby breaking the U(1) symmetry in their ground states. The in-plane angle of the spin density is analogous to the phase of the wave function in mass superfluid, and, accordingly, the supercurrent of spin (projected onto the symmetry axis) is proportional to the gradient of the in-plane angle. On the other hand, easy-axis magnets, which have the U(1) spin-rotational symmetry like the easy-plane ones, do not break the symmetry spontaneously because spins order along the symmetry axis in their two ground states. Easy-axis magnets, therefore, do not support spin superfluidity as a metastable state. Instead, they support a domain wall (DW)~\cite{*[][{, and references therein.}] KosevichPR1990}, which is a topological soliton smoothly connecting the two ground states. Since the two directions along the symmetry axis are the only directions that are invariant under spin rotations, a DW connecting those two states necessarily breaks the U(1) symmetry, which signals the possible existence of spin superfluidity in it.

\begin{figure}
\includegraphics[width=\columnwidth]{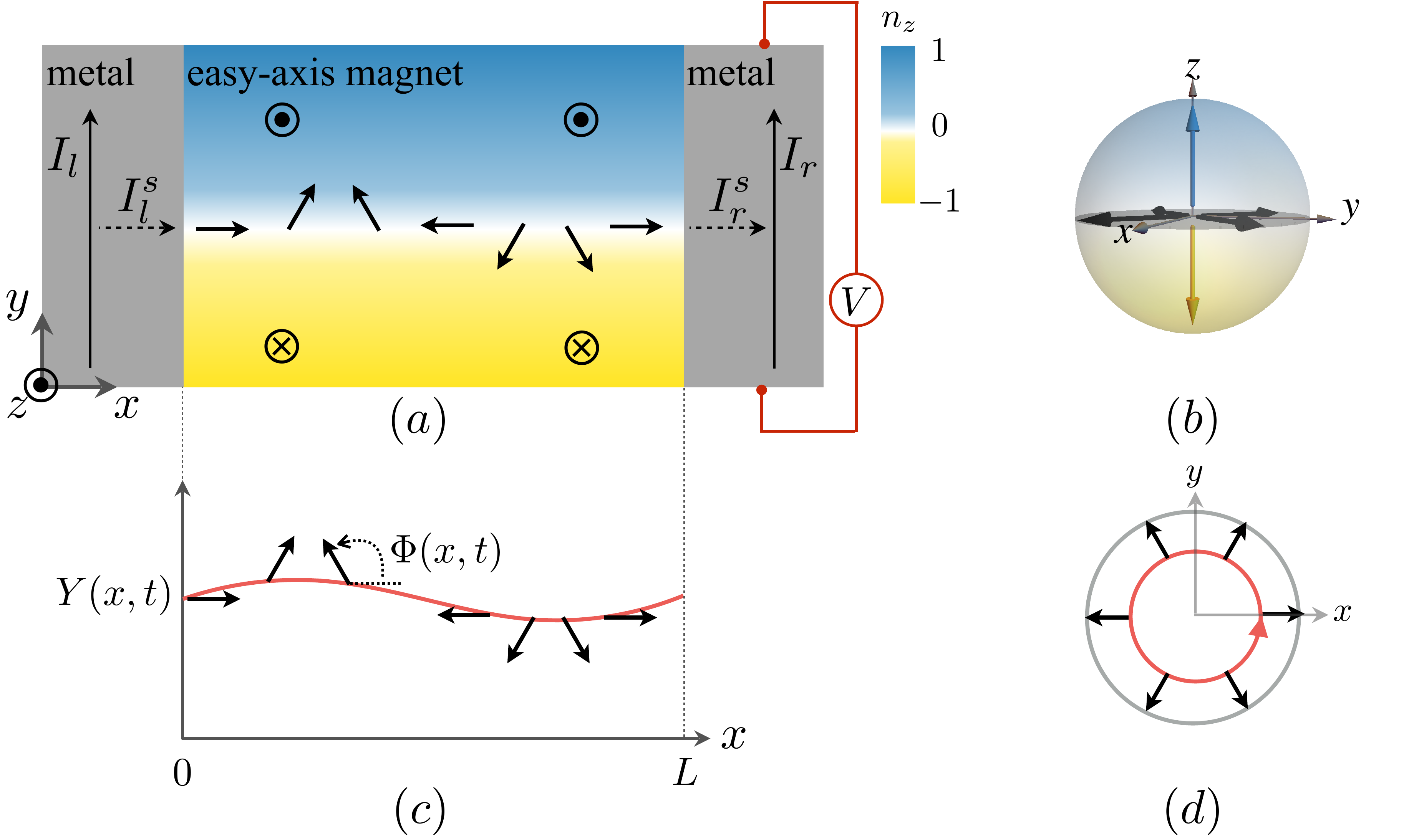}
\caption{(a) An illustration of a DW stretched along the $x$ direction, which is carrying a spin supercurrent. (b) Mapping of the spin texture in (a) onto the unit sphere, characterized by its skyrmion charge $Q = 1$. (c) An illustration for the simplified description of the DW using its vertical position $Y(x, t)$ and in-plane angle $\Phi(x, t)$. (d) Mapping of the DW state in (c) onto the unit circle, characterized by its winding number $w = 1$.}
\label{fig:fig1}
\end{figure}

Pursuing the idea, we study spin transport through a DW in a two-dimensional easy-axis magnet, the main results of which can be summarized as follows. First, we show that a DW indeed can serve as a conduit for spin superfluid. The spin supercurrent flowing through the DW is realized as a spiraling spin texture lying within the plane perpendicular to the easy axis. Secondly, we find the associated phase slips disturbing the spin supercurrent~\cite{*[][{, and references therein.}] HalperinIJMPB2010} are peculiar in that they accompany creation or destruction of skyrmions~\cite{SkyrmePRSA1961}, swirling spin textures wrapping the unit sphere once, as a result of the conservation of the total skyrmion charge~\cite{*[][{, and references therein.}] Garst2016}. See Fig.~\ref{fig:fig1} for schematic illustrations. The DW thus can provide a good tool for generating and detecting skyrmions. Thirdly, we show that the DW harbors the emergent magnetic flux for magnons along its length~\cite{KovalevEPL2012, *KongPRL2013, *IwasakiPRB2014, *SchuttePRB2014-2}, whose density is proportional to the spin supercurrent flowing through it. By engineering a periodic array of DWs, we can therefore create an emergent magnetic superlattice for magnons, which can be useful for utilizing magnons in spintronics analogously to the electronic counterparts~\cite{*[][{, and references therein.}] NogaretJPCM2010}. Lastly, magnons traveling across the DW can trigger the spin supercurrent by transferring their spin angular momentum to the DW, providing the magnonic control of the spin superfluid.

\emph{Spin superfluid in a DW.}|Our model system is a quasi-two-dimensional ferromagnet with easy-axis anisotropy~\footnote{We focus on an easy-axis ferromagnet for concreteness here, but it is straightforward to show that the case of an easy-axis antiferromagnet is closely analogous.}. For temperatures much below the Curie temperature, the low-energy dynamics of the magnet can be described by the direction of the local spin density, $\mathbf{n} \equiv \mathbf{s} / s$. The Hamiltonian is given by
\begin{equation}
\label{eq:H}
H = \int dxdy \left[ A (\boldsymbol{\nabla} \mathbf{n})^2 + K (1 - n_z^2) \right] / 2 \, ,
\end{equation}
where the positive coefficients $A$ and $K$ parametrize the stiffness of the order parameter and the strength of easy-axis anisotropy, respectively. The system respects the time-reversal symmetry and the U(1) symmetry with respect to the global spin rotations about the $z$ axis. In the two ground states, spins order along the $z$ direction, $\mathbf{n} \equiv \pm \hat{\mathbf{z}}$, by breaking the time reversal symmetry, but not the U(1) symmetry. In a continuous field theory, the discrete degeneracy of the ground states entails a DW interpolating them~\cite{*[][{, and references therein.}] Kardar}. The north and south poles on the unit sphere, which constitute the ground states, are the only points that are fixed under rotations about the $z$ axis. A DW connecting them, therefore, breaks the U(1) symmetry spontaneously, which can be exploited to realize spin superfluid~\cite{SoninAP2010}. See Fig.~\ref{fig:fig1}(a) for a schematic of the system with a DW carrying a spin supercurrent.

We consider a DW that is pinned around the straight line defined by $y = 0$. The pinning potential, which is omitted in Eq.~(\ref{eq:H}), can be engineered by, e.g., locally decreasing the magnitude of easy-axis anisotropy~\cite{BauerNN2013, *FrankenAPL2013, *FrankenJPCM2012}. Without loss of generality, we assume that the spins point at the north pole in the top and the south pole in the bottom, $\mathbf{n} \rightarrow \pm \hat{\mathbf{z}}$ as $y \rightarrow \pm \infty$. The vertical position of the DW is represented by $Y(x, t)$, at which the spin is in the $xy$ plane. The in-plane angle at the position $(x, Y)$ is denoted by $\Phi(x, t)$. The contribution from an infinitesimal segment of the DW to the total spin is given by $d  S_z = - 2 s Y(x, t) dx$; $Y$ thus represents the local spin density. Within the collective-coordinate approach~\cite{ThielePRL1973, *TretiakovPRL2008}, the low-energy dynamics of the DW can be described by the two fields $Y$ and $\Phi$ with the Hamiltonian~\footnote{Supplemental Material contains the derivations for the dynamics of the one-dimensional DW~\cite{SchryerJAP1974}, the calculations of phase-slip-induced voltages, the discussions on simulation results, and three videos showing the magnetization evolutions.}
\begin{equation}
\label{eq:H2}
H = \int dx (\kappa Y^2  + \eta \Phi'^2) / 2 \, ,
\end{equation}
where $\kappa$ represents the magnitude of the pinning potential, $\eta \equiv 2 \sqrt{A^3 / K}$ parametrizes the stiffness of the field $\Phi$, and $'$ is the spatial derivative. 

As we learn from quantum mechanics, the total spin projected onto the $z$ axis is the generator of the spin rotations about the same axis~\cite{LL3}, which yields the following Poisson bracket~\cite{Goldstein}: $\{ \phi (\mathbf{r}, t), s n_z (\mathbf{r}', t) \} = \delta(\mathbf{r} - \mathbf{r}')$, where $\phi$ is the azimuthal angle of $\mathbf{n}$. Within the collective-coordinate approach for the DW dynamics, this translates into
\begin{equation}
\label{eq:pb}
\left\{ Y(x, t), \Phi(x', t) \right\} = \delta (x - x') / 2 s \, ,
\end{equation}
which shows that the two fields are canonically conjugate. The equations of motion for them can be obtained from the Hamiltonian and the Poisson bracket:
\begin{subequations}
\label{eq:eom}
\begin{align}
2 s \dot{\Phi} &= - \kappa Y \, , \\
- 2 s \dot{Y} &= \eta \Phi''  \, .
\end{align}
\end{subequations}
The first equation is analogous to the Josephson relation in a one-dimensional mass superfluid between the phase $\Phi$ of the wave function and the mass density $\propto Y$; the second equation is the coarse-grained continuity equation of the spin~\cite{HalperinPR1969}, in which the left-hand side is the time evolution of the spin density integrated over the $y$ axis and the right-hand side is the (negative) divergence of the spin current along the DW, $I^s \equiv - \eta \Phi'$. The analogy between the dynamics of the DW and that of a mass superfluid leads us to conclude that the DW supports superfluid spin transport flowing through it~\cite{SoninAP2010}. Unlike the ideal dissipationless situation that we have considered heretofore, generic spin systems are subject to dissipation caused by, e.g., spin-lattice coupling. It can be effectively captured by adding the Gilbert damping terms to the equations of motion~\cite{Note2}:
\begin{subequations}
\label{eq:eom2}
\begin{align}
2s\dot{\Phi} &= - \kappa Y - 2 \alpha s \dot{Y} / \lambda \, , \\
- 2 s \dot{Y} &= \eta \Phi'' - 2 \alpha s \lambda \dot{\Phi} \, ,
\end{align}
\end{subequations}
where $\alpha$ is the Gilbert damping constant and $\lambda \equiv \sqrt{A / K}$ parametrizes the DW width.

Spin can be injected into or ejected from the magnet by sandwiching it with two metals that exhibit spin Hall effects~\cite{SinovaRMP2015} as shown in Fig.~\ref{fig:fig1}(a). In the presence of charge currents $I_l$ and $I_r$ flowing in the $y$ direction in the left and right metals, respectively, matching the spin current across the interface and that of the bulk leads to the following boundary conditions:
\begin{subequations}
\label{eq:bc}
\begin{align}
2 \lambda \left[ \vartheta I_l - \gamma \dot{\Phi} (x = 0, t) \right] &= - \eta \Phi' (x = 0, t) \, , \\
2 \lambda \left[ \vartheta I_r + \gamma \dot{\Phi} (x = L, t) \right] &= - \eta \Phi' (x = L, t) \, ,
\end{align}
\end{subequations}
where $L$ is the length of the magnet. In the first equation, $2 \lambda \vartheta I_l$ is the spin current injected from the left metal to the magnet, parametrized by the coefficient $\vartheta \equiv (\hbar / 2 e d_x) \tan \Theta$ with $d_x$ the normal-metal width in the $x$ direction and $\Theta$ the effective interfacial spin Hall angle~\cite{TserkovnyakPRB2014}; $2 \lambda \gamma \dot{\Phi}$ is the spin pumping from the magnet into the metal, parametrized by the coefficient $\gamma \equiv \hbar g^{\uparrow \downarrow} d_z / 4 \pi$ with $g^{\uparrow \downarrow}$ the effective interfacial spin-mixing conductance~\cite{TserkovnyakPRB2014} and $d_z$ the thickness of the interface in the $z$ direction; the right-hand side is the spin current in the magnet evaluated at the left interface. The equations~(\ref{eq:eom2}) for the dynamics of the DW in conjunction with the boundary conditions~(\ref{eq:bc}) are completely analogous to those for the dynamics of one-dimensional magnets with easy-plane anisotropy~\cite{SoninAP2010}. By adopting the known results for the latter system~\cite{TakeiPRL2014}, we obtain the solution for the steady state of our system, which precesses at the uniform frequency
\begin{equation}
\label{eq:omega}
\dot{\Phi} \equiv \Omega = \frac{\vartheta}{\gamma + \gamma_\alpha / 2} \frac{I_l - I_r}{2} 
\end{equation}
and carries the spin supercurrent
\begin{equation}
\label{eq:Is}
I^s = 2 \lambda \left[ \vartheta I_l - (\gamma + \alpha s x) \Omega \right] \, ,
\end{equation}
with $\gamma_\alpha \equiv \alpha s L$. These theoretical results for $\Omega$ and $I^s$ are confirmed by micromagnetic simulations performed with the aid of the software OOMMF~\cite{Donahue1999}. See Supplemental Material for the discussions on the simulation results~\cite{Note2}.

The superfluid spin transport through a DW can be probed experimentally in the following way proposed in Refs.~\cite{TakeiPRL2014}. The uniform spin precession at the frequency $\Omega$ induces the inverse spin Hall voltage in the metals, $\Delta V = \pm 2 \lambda \vartheta \Omega = \pm 2 \lambda \vartheta^2 (I_l - I_r) / [\gamma (1 + L / L_\alpha)]$, where the upper (lower) sign corresponds to the left (right) metal and $L_\alpha \equiv 2 \gamma / \alpha s$ is the crossover length. For a numerical estimate, let us consider the non-local generation of the voltage $\Delta V$ in the right metal due to the charge current $I_l$ in the left metal. We take the following material parameters for Pt$|$YIG$|$Pt compounds: $\lambda \sim 50$nm~\cite{JiangPRL2013}, $s \sim 10 \hbar / $nm$^3$~\cite{BhagatPSS1973}, $\alpha \sim 10^{-4}$, $\Theta \sim 0.1$, and $g^{\uparrow \downarrow} \sim 5/$nm$^2$~\cite{KajiwaraNature2010, *SandwegPRL2011, *HahnPRB2013}, which yields the crossover length $L_\alpha \sim 1\mu$m~\cite{TakeiPRL2014}. When using $d_x = 5$nm for the platinum geometry and  $J_l = 10^{10}$A/m$^2$ for the charge-current density~\cite{KajiwaraNature2010}, we obtain $|\Delta V| \sim 5 \times (1 + L / L_\alpha)^{-1}\mu$V. Observation of the algebraic dependence of $\Delta V$ on $L$ can serve as evidence for the superfluid spin transport.

\begin{figure}
\includegraphics[width=\columnwidth]{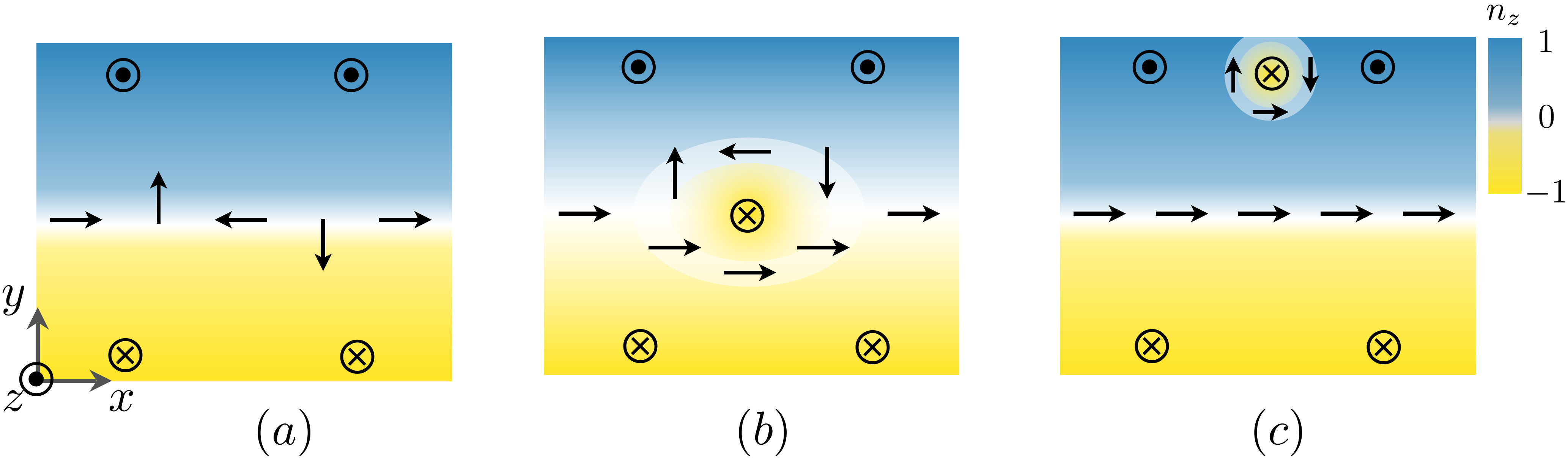}
\caption{A schematic illustration of a phase slip that decreases the winding number $w$ from $1$ in the initial state (a) to $0$ in the final state (c) via the intermediate state (b), in which $w$ is not well defined. (c) An isolated skyrmion with the skyrmion charge $Q = \Delta w = 1$ leaves the ferromagnet through the boundary.}
\label{fig:fig2}
\end{figure}

\emph{Phase slips creating skyrmions.}|One-dimensional superfluids are susceptible to dissipation; thermal and quantum fluctuations engender finite resistance disturbing the supercurrent via phase slips~\cite{HalperinIJMPB2010}, to which the spin supercurrent is not an exception~\cite{SoninAP2010, KimPRB2016, *KimPRL2016}. We will describe a topological aspect of the phase slips occurring to DW spin superfluid in the DW, which is absent in conventional superfluids. For the conceptual convenience of the discussion, we assume the periodic boundary conditions along the $x$ direction and the presence of one DW pinned at the straight line $y = 0$. Then the metastable state of the system can be classified by the integer U(1) winding number of the angle $\Phi$ along the DW,
\begin{equation}
w = \frac{1}{2 \pi} \int dx \, \partial_x \Phi \, .
\end{equation}
The spin supercurrent sustained by the metastable state is proportional to the winding number, $I^s = - 2 \pi \eta w / L$. Now, let us consider the spin texture expanded over the two-dimensional plane of the magnet. It covers the unit sphere integer number of times, which is referred to as the skyrmion charge~\cite{SkyrmePRSA1961, BelavinJETP1975}:
\begin{equation}
\label{eq:Q}
Q = \frac{1}{4 \pi} \int dxdy \, \mathbf{n} \cdot \partial_x \mathbf{n} \times \partial_y \mathbf{n} \, .
\end{equation}
For the metastable states, the skyrmion charge is equal to the winding number, $Q = w$. For example, in Figs.~\ref{fig:fig1}(a) and (b), the spin texture departs from the north pole of the unit sphere in the top of the magnet, covers the equator once along the DW (i.e., $w = 1$), and arrives at the south pole in the bottom. The spin texture covers the unit sphere exactly once, and thus it is classified by the skyrmion charge $Q = 1$. This relation between the winding number and the skyrmion charge has been found by \textcite{KudryavtsevN1998}.

If spins do not fluctuate, the metastable state with the nonzero winding number would be maintained indefinitely. Magnets, however, generally experience thermal or quantum fluctuations, which can drive transitions between metastable states by changing the winding number via phase slips. Due to the equivalence between the winding number and the total skyrmion charge of the metastable states, a phase slip should create or destroy a skyrmion with the skyrmion charge identical to the change of the winding number. As an example, let us consider the scenario illustrated schematically in Fig.~\ref{fig:fig2}, during which a particle-like skyrmion is produced. The initial metastable state has the winding number $w = 1$ and the skyrmion charge $Q = 1$. Via the phase slip, the winding number becomes zero $w = 0$, and the associated loss of the skyrmion charge drifts away from the DW and leaves the magnet through the boundary. 

\begin{figure}
\includegraphics[width=\columnwidth]{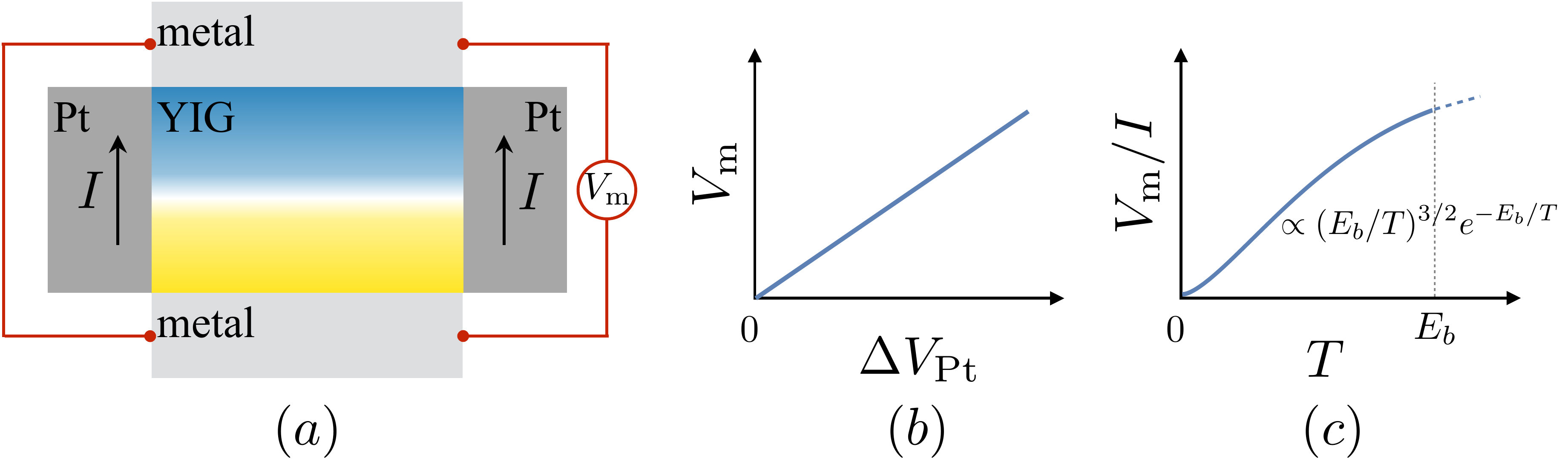}
\caption{(a) An experimental setup for probing phase slips. (b) The linear relationship between the two phase-slip-induced voltages, $\Delta V_\text{Pt}$ in the platinum and $V_\text{m}$ in the metals. (c) The temperature dependence of $V_\text{m} / I$ induced by thermally activated phase slips with the energy barrier $E_b$.}
\label{fig:fig4}
\end{figure}

Phase slips can be inferred by measuring the induced voltages in the proximate metals. See Fig.~\ref{fig:fig4}(a) for an experimental setup, in which two normal metals are attached to the Pt$|$YIG$|$Pt heterostructure along the length of the magnet and the charge current $I$ is flowing in two identical platinum. Phase slips unwind the spin texture with the average rate, which we denote by $\nu$; skyrmion charges are generated with the same frequency, $\dot{Q} = \nu$. The created skyrmions are drifted away from the DW via the Brownian motion~\cite{SchuttePRB2014, *KimPRB2015, *BarkerPRL2016} and eventually annihilated at the top and bottom boundaries, giving rise to the electromotive voltage in the metals, $V_\text{m} = P \hbar \nu / e$, where $e$ is the electric charge of electrons and $P$ is the dimensionless phenomenological parameter characterizing the degree of the magnetic proximity effect~\cite{VolovikJPC1987, *BarnesPRL2007, *TserkovnyakPRB2008, *TserkovnyakPRB2009}. In addition, phase slips result in the spin precession at the left and right boundaries, which generates the inverse spin Hall voltage in the platinum, $\Delta V_\text{Pt} = 2 \pi \lambda \vartheta \nu$~\cite{KimPRB2016}; this is analogous to the phase-slip-induced voltage in one-dimensional superconducting wires~\cite{HalperinIJMPB2010}. Two voltages, $V_\text{m}$ and $\Delta V_\text{Pt}$, are linearly proportional to each other as shown in Fig.~\ref{fig:fig4}(b) with the ratio independent of the temperature and the applied charge current, which can be tested against experiments. Based on a dimensional analysis, rough numerical estimates for $V_\text{m}$ and $\Delta V_\text{Pt}$ due to thermally activated phase slips can be obtained~\cite{Note2} by adopting the results for one-dimensional superconductors~\cite{HalperinIJMPB2010, McCumberPRB1970} and spin superfluids~\cite{KimPRB2016}. When using $K \sim 3 \times 10^{-6}$J/m$^2$~\cite{JiangPRL2013}, $T = 300$K, $L = 1$mm, $P \sim 1$ and $J = 10^{10}$A/m$^2$ for the charge-current density in addition to the  other parameters used above, we obtain $V_\text{m} \sim 7$nV and $\Delta V_\text{Pt} \sim 20$nV with the phase-slip energy barrier $E_b \sim 500$K. The functional dependence of $V_\text{m} / I$ on the temperature $T$ is shown in Fig.~\ref{fig:fig4}(c).

\begin{figure}
\includegraphics[width=\columnwidth]{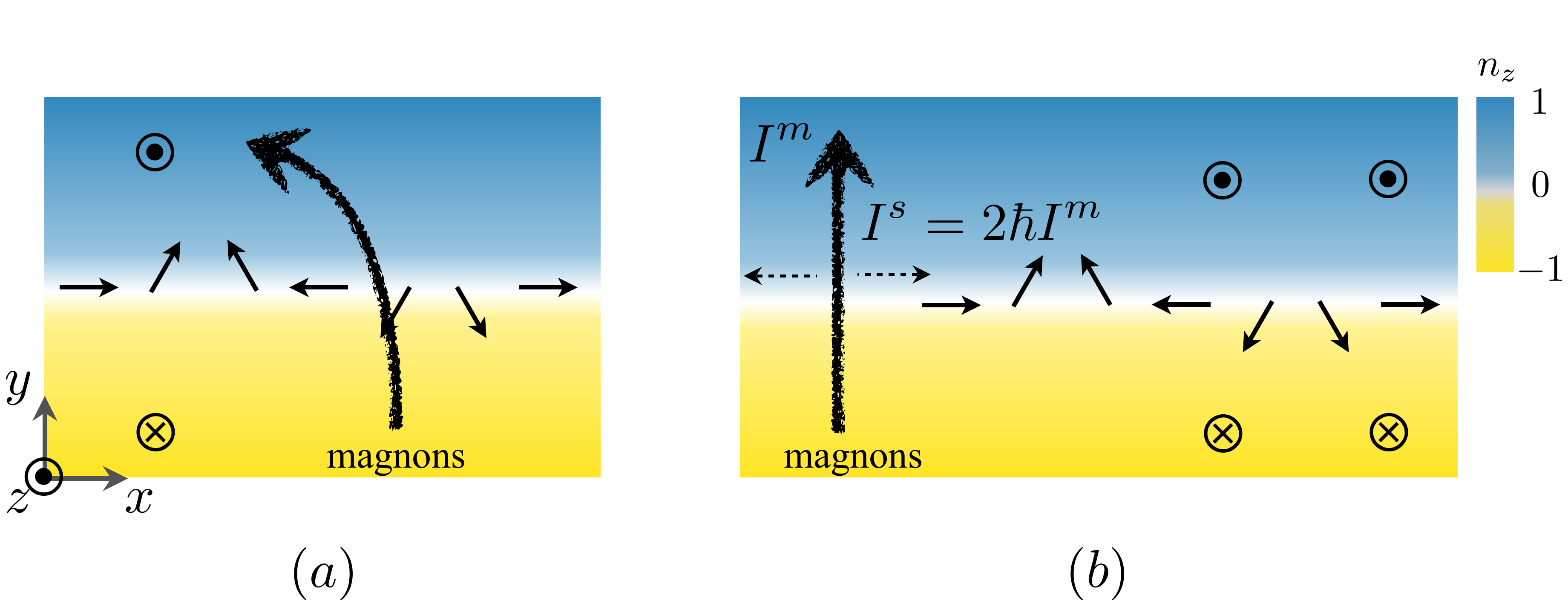}
\caption{(a) Magnons are deflected by the emergent magnetic field engendered by the spin supercurrent in the DW. (b) The magnon current $I^m$ flowing across the DW injects the spin supercurrent $I^s = 2 \hbar I^m$ into the wall via the magnonic spin-transfer torque.}
\label{fig:fig3}
\end{figure}

\emph{Interaction with magnons.}|In two-dimensional magnets, magnons are known to experience the emergent magnetic field proportional to the density of the skyrmion charge~\cite{KovalevEPL2012}, which can be harbored by the DW carrying a spin supercurrent. In the adiabatic limit, where a magnon keeps its spin antiparallel to the background spin texture, the emergent Lorentz force on it is given by
\begin{equation}
\mathbf{F} = \mathbf{v} \times b \hat{\mathbf{z}} \, ,
\end{equation}
where $\mathbf{v}$ is the velocity of the magnon and 
\begin{equation}
b = - \hbar \mathbf{n} \cdot \partial_x \mathbf{n} \times \partial_y \mathbf{n}
\end{equation}
is the strength of the emergent magnetic field, which is proportional to the integrand of the skyrmion charge Q~(\ref{eq:Q}). The emergent magnetic flux is localized at the DW, where a nontrivial spin texture exists. Let us consider a case shown in Fig.~\ref{fig:fig3}(a), where a magnon approaches the DW at the initial velocity $\mathbf{v}_i = v_0 \hat{\mathbf{y}}$. After traveling across the DW with the winding number $w$, the magnon acquires a finite $x$ component in its velocity:
\begin{equation}
v_{f, x} = \frac{1}{m} \int dt \, v_0 b= - \frac{4 \pi \hbar}{m} \frac{w}{L} \, ,
\end{equation}
to linear order in $b$, where $m \equiv \hbar s / 2 A$ is the effective mass of magnons. The DW carrying a spin supercurrent therefore gives rise to a magnon Hall effect, the strength of which can be controlled by manipulating the spin supercurrent.

In return, magnons can affect the spin texture via the magnonic spin-transfer torque~\cite{HinzkePRL2011, *YanPRL2011}. Let us consider a magnon current flowing in the positive $y$ direction as shown in Fig.~\ref{fig:fig3}(b). In the adiabatic regime, a magnon changes spin from $\hbar \hat{\mathbf{z}}$ to $- \hbar \hat{\mathbf{z}}$ while traveling across the DW. The spin-rotational symmetry of the system requires the conservation of spin angular momentum, and thus the DW should absorb $\Delta S_z = 2 \hbar \hat{\mathbf{z}}$ per each magnon passing through it. Accordingly, the magnon current $I^m$ across the DW will inject the spin current $I^s = 2 \hbar I^m$ into the DW, opening a possibility of magnonic manipulation of spin superfluid.

\emph{Discussion.}|The origin of the spin superfluid in a DW, a boundary between two different domains, has an instructive interpretation in terms of the topological property of magnons bands. In each domain, magnons' spin, which is antiparallel to the ground state, serve as a good quantum number because the U(1) spin-rotational symmetry is intact. In other words, the magnon band has a definite chirality, and it can be considered as a U(1)-symmetry-protected topological invariant of the band. Then, the spin superfluid between two domains with magnon bands of opposite chiralities has a natural interpretation as a gapless mode~\cite{KobayashiPRL2014, Garcia-SanchezPRL2015} between the two topologically distinct phases, analogous to the edge states of topological insulators~\cite{HasanRMP2010, *QiRMP2011}.

The U(1) spin-rotational symmetry is crucial for intact spin superfluidity, but it can be broken, e.g., by shape anisotropy or spin-orbit coupling, which will give rise to a critical barrier for the spin supercurrent to flow through the DW~\cite{SoninAP2010}. The barrier can be overcome by thermal fluctuations at finite temperatures, where thermally populated DWs can carry the spin current by mimicking superfluid spin transport~\cite{KimPRB2015-5}.

The production of particle-like skyrmions by the phase slips has been considered here as an example, whereas the more common scenario would be the generation of lumps that are not proper to be considered as particles. According to the Hamiltonian $H$~(\ref{eq:H}), they can decrease their radius without any energy cost and disappear by collapsing into a single lattice point~\cite{CaiPRB2012}. There are several ways to stabilize skyrmions in magnetic systems. For example, the interfacial Dzyaloshinskii-Moriya interaction~\cite{ThiavilleEPL2012, *YuNL2016}, which can be induced by proximate heavy metals, can stabilize skyrmions with a fixed radius.

\begin{acknowledgments}
We are grateful to Fenner Harper, Rahul Roy, and Oleg Tchernyshyov for insightful discussions. We also thank the anonymous referees, whose comments and questions led to the significant improvement of the Letter. This work was supported by the Army Research Office under Contract No. W911NF-14-1-0016.
\end{acknowledgments}

\bibliographystyle{/Users/evol/Dropbox/School/Research/apsrev4-1-nourl}
\bibliography{/Users/evol/Dropbox/School/Research/master}

\end{document}


\title{Supplemental Material: Magnetic Domain Walls as Hosts of Spin Superfluids and Generators of Skyrmions}

\author{Se Kwon Kim}
\author{Yaroslav Tserkovnyak}
\affiliation{Department of Physics and Astronomy, University of California, Los Angeles, California 90095, USA}

\date{\today}

\maketitle

In this Supplemental Material, the dynamics of the domain wall will be derived within the collective coordinate approach and the numerical estimates for phase-slip-induced voltages are calculated. Also, the micromagnetic simulation results for confirming superfluid spin transport within the domain wall are presented.

\section{The domain-wall dynamics}

To make the spin-rotational invariance more explicit, it is convenient to parametrize the unit vector with spherical angles: $\mathbf{n} = (\sin \theta \cos \phi, \sin \theta \sin \phi, \cos \theta)$. The dynamics of the system can be described by the Lagrangian, $L = \int dV s \cos \theta \dot{\phi} - H$. The set of the three parameters, $A, K$, and $s$, in the Lagrangian defines the natural units of energy, length, and time of the problem, which are given by $\epsilon_0 = A$, $\lambda = \sqrt{A / K}$, and $t_0 = s / K$, respectively. By solving the Landau-Lifshitz equation associated with the Lagrangian, we can obtain a solution for a static domain wall lying along the $x$ axis, which is given by
\begin{equation}
\label{eq:dw}
\cos \theta(x, y) = \tanh[(y - Y) / \lambda] \, , \quad \phi(x, y) \equiv \Phi \, ,
\end{equation}
with $\sin \theta (x, y) = \sech[(y - Y) / \lambda]$~[37]. Here, $Y$ and $\Phi$ represent the center position of the domain wall in the $y$ direction and the azimuthal angle therein. The Goldstone modes associated with the breaking of these continuous symmetries can be accounted for by promoting the two domain-wall parameters, $Y$ and $\Phi$, to dynamic variables, $Y(x, t)$ and $\Phi(x, t)$~[31]. Since all the other modes have finite excitation gaps proportional to the natural unit of energy~[32], we can describe the low-energy dynamics of the magnet with these two dynamic fields. 

By plugging the domain-wall solution in Eq.~(\ref{eq:dw}) into the Lagrangian, we obtain 
\begin{equation}
L = \int dx \left[ - 2 s Y \dot{\Phi} - (\tau Y'^2 + \eta \Phi'^2 + \kappa Y^2) / 2 \right] \, ,
\end{equation}
where $'$ and $\dot{}$ represent the derivatives with respect to the spatial $x$ and the temporal coordinate $t$, respectively. Here, the first term describes the gyrotropic coupling between the position $Y$ and the angle $\Phi$ of the domain wall, which yields the Poisson bracket~(3) between them; the second term describes the tension of the domain wall with $\tau \equiv 2 \epsilon_0 / l_0$; the third term represents the stiffness for the angle change with $\eta \equiv 2 \epsilon_0 l_0$; the last term with the dimensionless parameter $\kappa$ represents a pinning potential for the domain wall. The pinning potential makes the spin-wave dispersions linear $\omega \propto k$, which is required to realize spin superfluid~[2]. We ignore the tension of the domain wall here by focusing on the dynamics with sufficiently long wavelengths. The dissipative dynamics can be effectively captured in the Lagrangian formalism by the Rayleigh dissipation function, $R = \alpha s \int dV \dot{\mathbf{n}}^2 / 2$, which represents the half of the energy dissipation rate~[16]. Plugging the domain-wall solution into it yields $R = \alpha s \lambda \int dx (\dot{Y}^2 / \lambda^2 + \dot{\Phi}^2)$. From the Lagrangian and the Rayleigh dissipation function, the equations of motion~(5) can be obtained. In the main text, the alternative Hamiltonian formalism is used, which can be obtained from the Lagrangian formalism developed here.

The spin current at the left and the right interfaces are given by
\begin{subequations}
\begin{align}
\sin^2 \theta ( \vartheta I_l - \gamma \dot{\phi} ) &= - A \sin^2 \theta \partial_x \phi \, ,  \\
\sin^2 \theta ( \vartheta I_r + \gamma \dot{\phi} ) &= - A \sin^2 \theta \partial_x \phi  \, ,
\end{align}
\end{subequations}
respectively. By integrating the equations over the $y$ axis with the domain-wall solution, the boundary conditions in Eq.~(6) in terms of the collective coordinates $Y$ and $\Phi$ are obtained.

\section{Phase-slip-induced voltages}

Numerical estimates for $V_\text{m}$ and $\Delta V_\text{Pt}$ induced by thermally activated phase slips can be calculated as follows. The thorough analysis of phase slips is beyond the scope of the present work, and we will content ourselves with a simple dimensional analysis by adopting the results obtained for one-dimensional superconductors~[5, 29] and spin superfluids~[24]. We are assuming here that the length scale introduced by the pinning potential is larger than the domain-wall width $\lambda = l_0$ so that the domain-wall profile is not distorted significantly. At a temperature $T$, the decaying rate $\nu$ of the spin supercurrent is given by 
\begin{equation}
\nu \sim \frac{\alpha}{t_0} \frac{L}{l_0} \frac{I^s}{\epsilon_0} \left( \frac{E_b}{T} \right)^{3/2} e^{-E_b / T} \, ,
\end{equation}
where $E_b \sim \epsilon_0$ represents the energy barrier for phase slips and $I_s = 2 \lambda \vartheta I$ [Eq.~(8)] is the spin supercurrent. Here, $\alpha/t_0$ is the inverse of the time scale for the relaxation dynamics of the magnet; $L / l_0$ represents the number of possible independent phase-slip locations; $I^s / \epsilon_0$ is the magnitude of the spin supercurrent in units of the characteristic energy scale; $E_b / T$ is the energy barrier in units of the temperature. See Refs.~[24] for more detailed discussions. When using $K \sim 3 \times 10^{-6}$J/m$^2$~[21], $T = 300$K, $L = 1$mm, $I / d_x d_z = 10^{10}$A/m$^2$, and $P \sim 1$ in addition to the parameters for Pt$|$YIG$|$Pt compounds used in the main text, we obtain $V_\text{m} \sim 7$nV and $\Delta V_\text{Pt} \sim 20$nV.

\section{Micromagnetic simulations}

\begin{figure}
\includegraphics[width=\columnwidth]{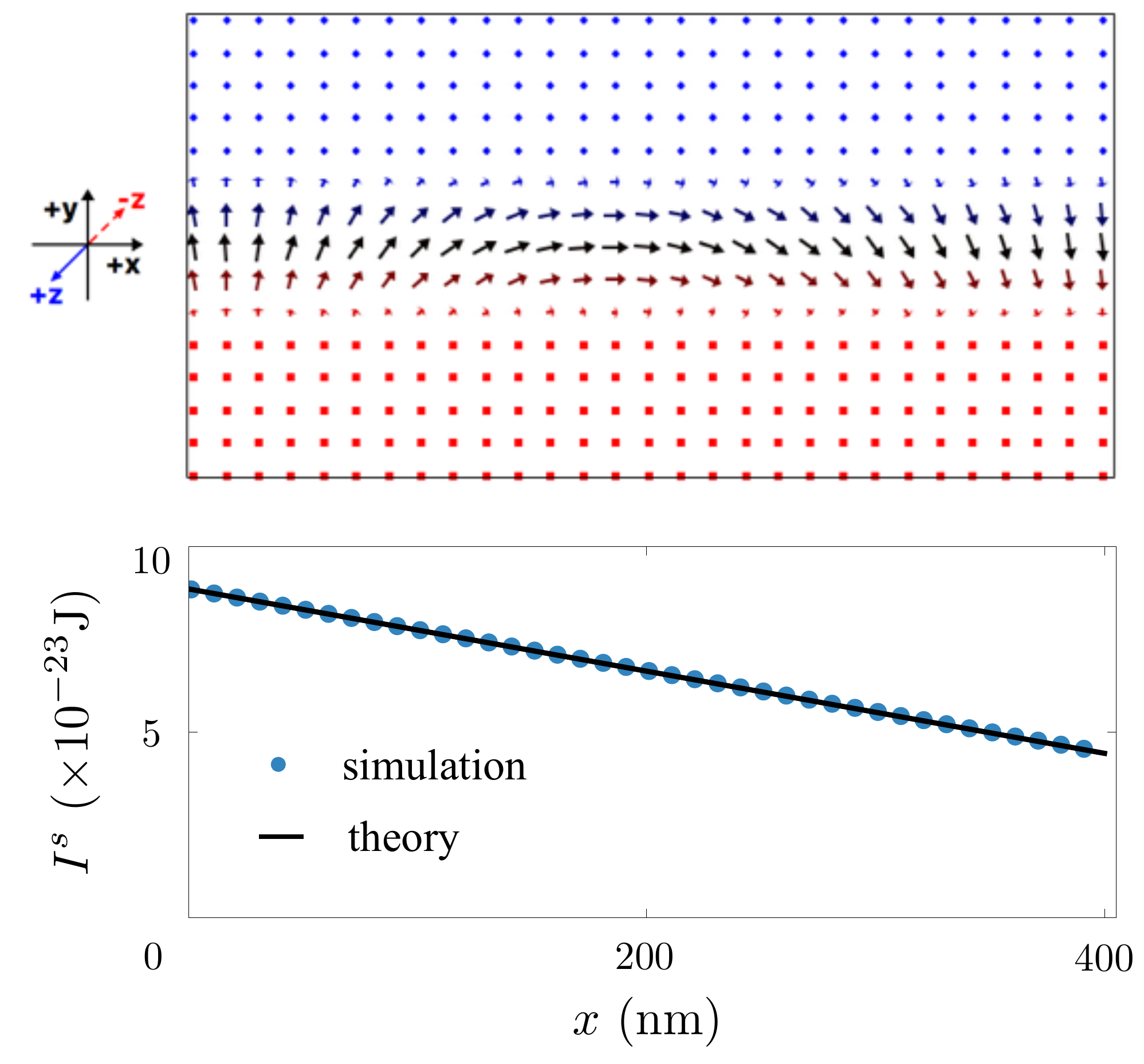}
\caption{The upper figure shows the magnetization carrying a spin supercurrent within the domain wall. The dimensions of the magnet are given by $L_x = 400$nm and $L_y = 200$nm; the electric current densities at the left and right boundaries, which we used to induce the spin-transfer torque in OOMMF, are given by $J_l = 4 \times 10^9$A/m$^2$ and $J_r = 2 \times 10^9$A/m$^2$, respectively. The lower figure shows the spin current $I_s(x)$ flowing across the cross section in the $yz$ plane. The theory [Eq.~(\ref{eq:Is-n})] explains the simulation results well. See the text for further discussions.}
\label{fig:sfig1}
\end{figure}

We performed numerical simulations for the Hamiltonian $H$ [Eq.~(1)] with the aid of the micromagnetic solver OOMMF~[20] in order to confirm our theory of superfluid spin transport through a magnetic domain wall. Specifically, we verified the results for the global precession frequency $\Omega$ [Eq.~(7)] and the spatial distribution of the spin current $I^s$ [Eq.~(8)] in the steady states. For the geometry of the film, we used the thickness $L_z = 2$ nm and the width $L_y = 200$nm; we have considered two lengths $L_x = 400$ and $800$nm to study its effects on the superfluid spin transport. By adopting the material parameters of YIG reported in Ref.~[21] with modifications to facilitate the simulation, we used the following values in simulations: the saturation magnetization $M_n = 8.4 \times 10^4$ A/m, the gyromagnetic ratio $\gamma_n = 1.75 \times 10^{11}$ s$\cdot$A/kg, the 2D easy-axis anisotropy coefficient $K_n = 2.4 \times 10^{-6}$ J/m$^2$, the 2D stiffness coefficient $A_n = 2.4 \times 10^{-22}$ J, the lattice constant $a_n = 2$nm, and the Gilbert damping constant $\alpha_n = 0.1$. These yield the domain wall width of $\lambda_n = \sqrt{A_n / K_n} = 10$nm that is smaller than the observed width of $50$ nm~[21]. To realize a pinning potential, we used the following spatially-modulated anisotropy, $K_n(y) = K_n [ 1 - 2 \sech(y / \lambda_n)]$, which confines the domain wall around the straight line given by $y = 0$. 

To perform numerical simulations, we first obtained the equilibrium states by minimizing the energy with OOMMF. We observed that the spatially-varying anisotropy yields an effective domain-wall width, $\lambda_n^* \equiv \int dy \sin^2 \theta / 2 = 17$nm, which is broader than the bare value for the uniform anisotropy, $\lambda_n = 10$nm. We then performed the dynamic simulations for $500$ns, which is long enough to reach the steady states. The magnetostatic energy is not included in the simulations as it is absent in the Hamiltonian $H$ [Eq.~(1)].

\begin{figure}
\includegraphics[width=0.9\columnwidth]{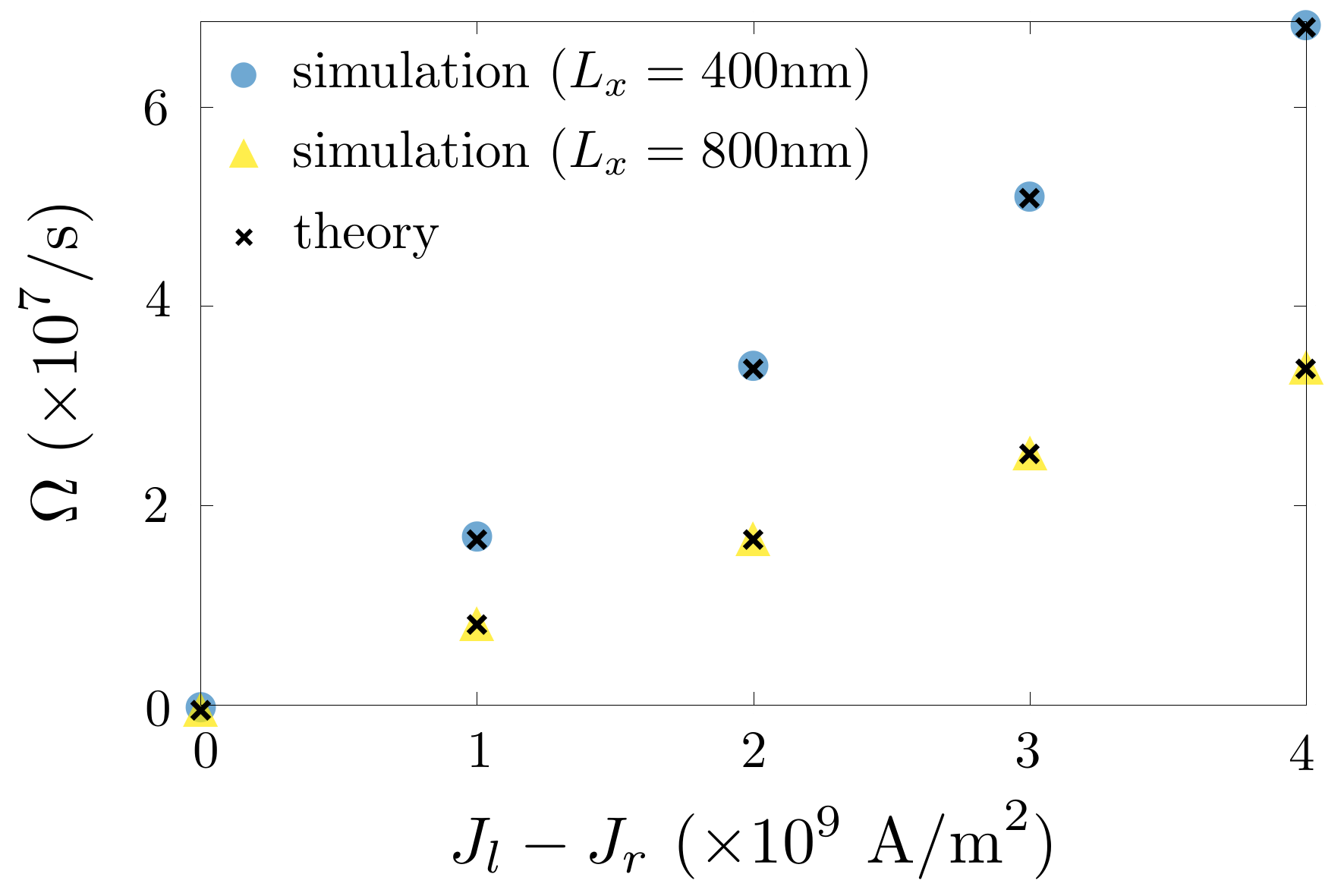}
\caption{The precession frequency, $\Omega$, as a function of the difference between two electric current densities, $J_l - J_r$. The theory [Eq.~(\ref{eq:omega-n})] explains the simulation results well. See the text for the further discussions.}
\label{fig:sfig2}
\end{figure}

The spin-transfer torque induced by the heavy metals at the interfaces (that we considered in the main text) can be simulated by the spin-transfer torque option in OOMMF, which models the spin transfer from proximate spin-polarized metals via an electric current. In the main text, the spin-transfer torque per unit length is given by $\boldsymbol{\tau} = \pm \vartheta I \mathbf{n} \times (\hat{\mathbf{z}} \times \mathbf{n})$ with the above (below) sign for the left (right) interface; the factor $\vartheta I$ appears in the expression for $\Omega$ [Eq.~(7)] and $I_s$ [Eq.~(8)]. In OOMMF, the corresponding spin-transfer torque per unit length (with the additional parameters set by $P = 1$ and $\Lambda = 1$) is given by $\boldsymbol{\tau} = (\hbar J d_x / 2 e) \mathbf{n} \times [(\pm \hat{\mathbf{z}}) \times \mathbf{n}]$, where $J$ is the current density flowing perpendicular to the film, $d_x = a$ is the width of the spin-polarized metals in the $x$ direction, $e > 0$ is the electric charge of an electron, and $\pm \hat{\mathbf{z}}$ is the spin-polarization direction of the metals (the above sign for the left metal and the below sign for the right metal). By matching the two expressions for the spin-transfer torque, we can identify $\hbar J d_x / 2 e$ as $\vartheta I$. We used $J_l = 4 \times 10^{9}$ A/m$^2$ for the current density of the left metal, which amounts to $\vartheta I_l = 2.6 \times 10^{-15}$ J/m. We varied the current density of the right metal, $J_r = 0, 1, 2, 3, 4 \times 10^{9}$ A/m$^2$. The spin pumping from the magnet to the metals is not implemented in OOMMF, and thus the parameter $\gamma$ in the expressions for $\Omega$ [Eq.~(7)] and $I_s$ [Eq.~(8)] is zero. In terms of the parameters used in the simulations, the analytical expressions tested against the numerical results are given by
\begin{eqnarray}
I^s &=& \frac{\hbar d_x \lambda_n^*}{e L_x} \left[ (L_x - x) J_l + x J_r \right] \, , \label{eq:Is-n} \\
\Omega &=& \frac{\gamma_n}{\alpha M_s L_x} \frac{\hbar}{2 e} (J_l - J_r) \, \label{eq:omega-n} .
\end{eqnarray}
These results, which are obtained with $\gamma \rightarrow 0$, correspond to the cases when the bulk damping $\gamma_\alpha$ exceeds the interfacial damping $\gamma$. In this limit, there is no need to include the spin pumping at the interfaces, which is compatible with OOMMF.

Let us discuss the simulation results. First, Fig.~\ref{fig:sfig1} shows a snapshot of the magnetization in the steady state, where the length of the magnet is $L_x = 400$nm and the current densities are $J_l = 4 \times 10^9$ A/m$^2$ and $J_r = 2 \times 10^9$ A/m$^2$. The observed in-plane spiraling texture of the magnetization within the domain wall represents a flow of a spin supercurrent~[2]. From the simulation results, we calculated the spin current as a function of the spatial coordinate $x$, $I^s_n (x) = - A_n \int dy \, \sin^2 \theta \, \partial_x \phi$. From Fig.~\ref{fig:sfig1}, we can see that the theoretical result for $I^s$ [Eq.~(\ref{eq:Is-n})] agrees well with the simulation results $I^s_n$. The spin current decreases linearly as a function of $x$ because of the spin damping. Secondly, Fig.~\ref{fig:sfig2} shows the precession frequency $\Omega$ as a function of the current difference, $J_l - J_r$, for two lengths of the magnet, $L_x = 400$ and $800$nm. The simulation results for $\Omega$ were obtained by calculating the average time derivative of the azimuthal angle of the total in-plane magnetization with the magnetization between $300$ns and $500$ns, when the system is in the steady states. We can see that the theoretical result in Eq.~(\ref{eq:omega-n}) explains the simulation results well. The precession frequency is smaller for the longer length because the total dissipation due to the damping is proportional to the volume of the magnet. We provide three videos made from the snapshots of the magnetization: video1.mp4 for $L_x = 400$nm and $J_r = 2 \times 10^9$ A/m$^2$, video2.mp4 for $L_x = 800$nm and $J_r = 0$ A/m$^2$, and video3.mp4 for $L_x = 800$nm and $J_r = 4 \times 10^9$ A/m$^2$. In all the videos, the finite-current density, $J_l = 4 \times 10^9$ A/m$^2$, is maintained at the left.